\documentclass[twocolumn]{emulateapj}

\bibpunct{(}{)}{;}{a}{}{,}

\newcommand{\msun}{\mbox{$M_\odot$}}
\newcommand{\rsun}{\mbox{$R_\odot$}}
\def\be{\begin{eqnarray}}
\def\ee{\end{eqnarray}}
\def\lsim{\mathrel{\rlap{\lower3pt\hbox{\hskip1pt$\sim$}}
     \raise1pt\hbox{$<$}}} %less than or approx. symbol
\def\gsim{\mathrel{\rlap{\lower3pt\hbox{\hskip1pt$\sim$}}
     \raise1pt\hbox{$>$}}} %greater than or approx. symbol

\shorttitle{LMC X-3 as a relic of GRBs}
\shortauthors{Brown, Lee, Moreno M\'endez}

\begin{document}

\title{LMC X-3 May Be a Relic of a GRB Similar to Cosmological GRBs}

%\author
%{Gerald E. Brown,$^{1}$ Chang-Hwan Lee$^{2\ast}$ and Enrique Moreno M\'endez$^{1}$\\
%\normalsize{$^{1}$Department of Physics and Astronomy, State University of New York,}\\
%\normalsize{Stony Brook, New York 11794, USA}\\
%\normalsize{$^{2}$Department of Physics, Pusan National University, Busan 609-735, Korea}\\
%\normalsize{$^\ast$To C.-H. Lee correspondence should be addressed; E-mail: clee@pusan.ac.kr.} }

\author{Gerald E. Brown}
\affil{Department of Physics and Astronomy,
               State University of New York, Stony Brook, NY 11794, USA.}
\email{GEB: gbrown@insti.physics.sunysb.edu}

\author{Chang-Hwan Lee}
\affil{Department of Physics, Pusan National University,
              Busan 609-735, Korea.}
\email{CHL: clee@pusan.ac.kr}

\and

\author{Enrique Moreno M\'endez}
\affil{Department of Physics and Astronomy,
               State University of New York, Stony Brook, NY 11794, USA.}

\email{EMM: moreno@grad.physics.sunysb.edu}

%---------------------------------------------------------------------

\begin{abstract}

The present scenario for high-luminosity long $\gamma$-ray bursts is strongly influenced by the paper of Fruchter et al. (2006). Whereas the main contention of this paper that these GRBs occur in low-metallicity irregular galaxies is based on a considerable collection of observational results and although the main thesis is doubtless correct, the paper does not explain the dynamics that produces such GRBs and much of the discussion not directly concerning the main thesis is wrong.  We propose a dynamics and elucidate how the \citet{Fruchter06} results may be tested, in our neighborhood in the LMC, suggesting that LMC X-3 is a relic of a high luminosity explosion, probably accompanied by a GRB and hypernova explosion.  The way to test our suggestion is to measure the system velocity of the present black hole.  We correct errors of the Fruchter et al. paper in stellar evolution, so that the study of GRBs is consistent with it.  We show that the subluminous GRB 060218 had a low-mass black hole as central engine.

\end{abstract}


%--------------------------------------------------------------------
\section{Introduction}\label{Intro}

\citet{Fruchter06} collect an important amount of data on long $\gamma$-ray bursts.  They propose that the long-duration $\gamma$-ray bursts are associated with the most extremely massive stars and that they may be restricted to galaxies of limited chemical evolution.  Also that long $\gamma$-ray bursts are relatively rare in galaxies such as our own Milky Way.

In recent papers \citep{Mor07, BLMM07} we have developed the Blandford-Znajek model of GRBs into quantitative calculations of the angular momentum energy that can be delivered for GRBs and hypernova explosions.  This was possible because \citet{Lee02} showed how to work out the Kerr parameters of the rotating black holes.  We showed that the Galactic soft X-ray transient sources were relics of Galactic explosions and constructed the energy of all 15 of the known Galactic sources.

It turns out that most of the Galactic sources underwent subluminous  GRBs, not because they did not possess sufficient rotational energy, but mostly because the rotational energy was so large that it destroyed the accretion disk so quickly that the central engine was dismantled before the GRB could properly develop.  We give as examples the transient sources Nova Sco (GRO J1655$-$40) and Il Lupi (4U 1543$-$47) for which \citet{Lee02} had predicted Kerr parameters of $a_\star=0.8$ and which were measured, by \citet{Sha06}, at the present time, to be $a_\star=0.65-0.75$ and $a_\star=0.75-0.85$, respectively.  The natal rotational energy was $430$ bethes (one bethe$=10^{51}$ ergs) and the final (measured) energy is indistinguishable, within errors, from the natal energy.  \citet{Bro00} had reconstructed the GRB and hypernova explosion for Nova Sco.  The nature of the explosion could be reconstructed from the donor, which accepted a number of $\alpha$-particle nuclei, especially $^{32}$S which is special for hypernova explosions, but rare in supernova explosions.

Thus, from the near equality of the natal and present rotational energies, only a few percent of the available rotational energy could have been accepted in the GRB and hypernova explosion.

\citet{Mor07} and \citet{BLMM07} showed from population syntheses that the soft X-ray transient sources were sufficient in number to account for all of the subluminous bursts in our neighborhood.

What about the cosmological bursts which are the long $\gamma$-ray bursts considered by \citet{Fruchter06}?  The Woosley Collapsar model was invented to describe these.  In terms of numbers these are only a few percent of the subluminous bursts \citep{Lia07}.  We will not take issue with the Woosley Collapsar model describing these, because our binary evolution begins by the donor spinning the He star, progenitor of the black hole, to whatever angular momentum is needed. Then the donor decouples, acting only as a witness to the explosion, with the He star collapsing into a black hole in the same way as in the Woosley Collapsar model, the tidal locking between the donor and the He star being transferred to one between the donor and the black hole.  From the latter, \citet{Lee02} predicted the Kerr parameter of the black hole.

The main point developed by \citet{BLMM07} was that the rotational energy of the binary is roughly inversely proportional to the mass of the donor. This follows from Kepler's law and from the fact that in Case C mass transfer (following the He burning), the initial $a_i$s (distance between the giant and the companion) of the binaries are roughly equal.  The fact that, according to \citet{Fruchter06}, the long $\gamma$-ray bursts are in low-metallicity galaxies, does not elucidate the dynamics which produce the long bursts.  The dynamics result from the fact that low-metallicity galaxies tend to have stars of higher mass than Galactic.  The higher mass of the donors slows the binary down sufficiently that the rotational energy can be accepted by the central engine.  In other words, the question of energy is a ``Goldilocks" one.  It must be not too much, because in that case the central engine will be dismantled, and not too little, because that would only be sufficient for a subluminous burst, but for a long high luminosity $\gamma$-ray burst it must be just right.

We have learned enough about Kerr parameters from our calculations and from the Smithsonian-Harvard measurements to construct a ``guesstimate".  Namely, we believe that LMC X$-$3 underwent the closest (in energy) explosion in our Galaxy, to a cosmological GRB.  (Of course, one can say that LMC X$-$3 is not in our Galaxy, but in the LMC.)

We believe that in its $\sim1/3$ solar metallicity \citep{Russ90}, it tends towards the low-metallicity stars considered by \citet{Fruchter06}, so that at the least it has somewhere between Galactic and the low metallicity favored there.  There is uncertainty in the masses, but \citet{Davis06} have a value of $a_\star\simeq0.26$ for the present Kerr parameter.  They took $7\msun$ as the mass of the black hole, which would imply a donor of $\sim4\msun$ from the lower end of the measurements of \citet{Cow83}.  We take these to be representative; other investigators have found other masses, so we suggest our evolution as only a possible one.

All of the binaries in \citet{BLMM07} and \citet{Mor07} which had so much energy that they dismantled the black hole accretion disk had donor masses of $(1-2)\msun$, so the donor in LMC X$-$3 is at least double those masses.  The donor mass is close to the $\sim5\msun$ that \citet{Mor07} estimated would give the energy of a cosmological GRB.  In any case, LMC X$-$3 is the closest ``nearby" relic of binaries similar to the progenitors of cosmological GRBs.  We can, therefore, use it as an example to try to reconstruct the explosion and model the energy of the explosion.  We estimate the mass loss in the explosion, finding it to be substantial.  It is likely that the estimated system velocity can be measured, at least the radial component of it, which should test our prediction.

In this paper we wish to also summarize results of earlier calculations which have a direct bearing on the 5 papers in the 31 August 2006 Nature \citep{N0,N1,N2,N3,N4}, in order to show that useful evolutions of black holes have been carried out in the past, of which the 119 astronomers who signed these articles were unaware.  We show that these previous calculations have a direct bearing on the measurements of GRB 060218/SN 2006aj; namely that the central engine was a black hole, not the magnetar conjectured by most of the authors, and that the black hole was one of minimum black hole mass with an $\sim(18-20)\msun$ ZAMS (Zero Age Main Sequence) progenitor.  We also correct a number of errors in Galactic black hole evolution in the \citet{Fruchter06} article.

%--------------------------------------------------------------------
\section{Evolution of Black Holes in Our Galaxy}\label{GBHEvo}

We begin by expanding on the evolutionary discussions in \citet{Bro01}.  The history of black hole evolution in binaries (which is the only place where black holes could be studied in detail) was that whatever mass, within reasonable limits, one proposed, the binary would turn out to have had a neutron star rather than black hole as compact object.  The first clear explanation of this was given by \citet{Bro01}; namely, the evolution of black holes in our Galaxy depends upon binarity.  Namely, in Case A or B mass transfer (mass transfer while the giant star is in main sequence or red giant stage) the strong winds in our Galaxy blow off sufficient of the ``naked" He envelope so that the remaining core of metals was too light to evolve into a black hole; rather, it would end up as a neutron star.

Only in Case C mass transfer, if mass transfer following He burning were carried out, would the remaining core have the possibility of evolving into a black hole.  Now, just what the limit is for the lowest ZAMS mass star that will evolve into a black hole is determined by what we call the Woosley Ansatz.  In our opinion this is one of the most powerful developments in stellar evolution.  We will combine this ``Ansatz" with the \citet{BBAL} considerations of entropy in the Fe core.

The Woosley Ansatz basically divides the burning of $^{12}$C into low energy, $T\sim20$keV burning through the $^{12}$C$+\alpha\rightarrow^{16}$O process and the $T\sim80$keV $^{12}$C$+^{12}$C$\rightarrow^{24}$Mg etc. process.  The $^{12}$C is produced by $\alpha+^8$Be$^\star\rightarrow^{12}$C; i.e., essentially through $\alpha+\alpha+\alpha\rightarrow^{12}$C.  This is a three body process, going as the square of the density, $\rho^2$. The $^{12}$C is burned into $^{16}$O by the two body process $^{12}$C$+\alpha\rightarrow^{16}$O$+\gamma$, which goes as $\rho$.  As long as $^{12}$C is present, the latter reaction will take place.
However, with increasing $M_{\rm ZAMS}$, the density decreases. The entropy, which goes inversely with the density, is known to increase with $M_{\rm ZAMS}$. Therefore, there will be a value of $M_{\rm ZAMS}$ where the $^{12}$C is removed by the $^{12}$C$+\alpha\rightarrow^{16}$O as rapidly as it is formed by the $\alpha+\alpha+\alpha$ process.  At this value of $M_{\rm ZAMS}$ the $^{12}$C$+^{12}$C$\rightarrow^{24}$Mg etc. shuts off, because there is no $^{12}$C; actually, it shuts off once the central carbon abundance is less than $\sim15\%$ because there is not enough carbon  for convective (steady) burning.  At this point the burning processes are all of low temperature, $T\sim20$keV.  At this point the metallicity is close to zero, independent of the average metallicity of the star.

Now, this temperature is too low for neutrino-pairs to carry off appreciable energy and entropy (the relativistic $\nu$, $\bar{\nu}$ pair cross section goes as $T^{11}$ power), whereas copious amounts of entropy were carried off by $^{12}$C$+^{12}$C$\rightarrow^{24}$Mg etc.  What happens to the entropy which increases with increasing $M_{\rm ZAMS}$?

\citet{BBAL} showed that the entropy per nucleon in the Fe core of a star in advanced burning was $\sim1$ in units of $k_B$.  The only place the increase in the entropy can go is into an increase in the number of nucleons in the Fe core, once the burning is confined to the low-temperature ($T\sim20$keV) region.  Thus at the ZAMS mass at which the two-body process takes over completely from the three-body process, central abundance of $^{12}$C has decreased below $15\%$ and the increase in entropy comes from the Fe cores increasing rapidly with $M_{\rm ZAMS}$.

\begin{figure}
%\centerline{
\includegraphics[width=0.65\textwidth]{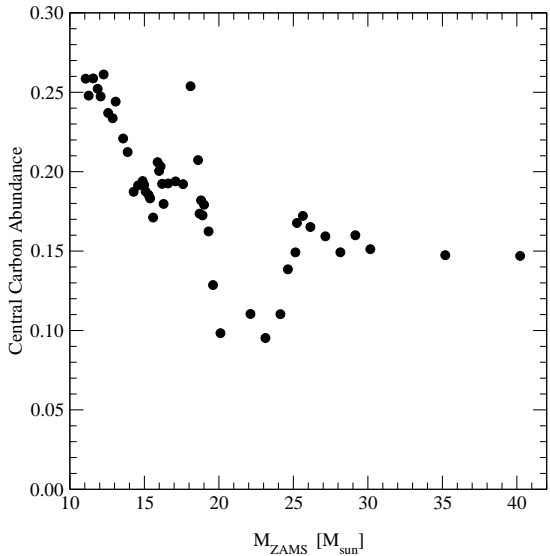}
%}
\caption{Central carbon abundance at the end of He-core burning for ``clothed" (single) stars as function of ZAMS mass.  The rapid drop in the central carbon abundance at ZAMS mass $M_{\rm ZAMS}\sim20\msun$ signals the disappearance of convective carbon burning. The resulting iron core masses are summarized in Fig.~\ref{fig:iron}.}
\label{fig:carbon}
\end{figure}

\begin{figure}
\centerline{\includegraphics[width=0.5\textwidth]{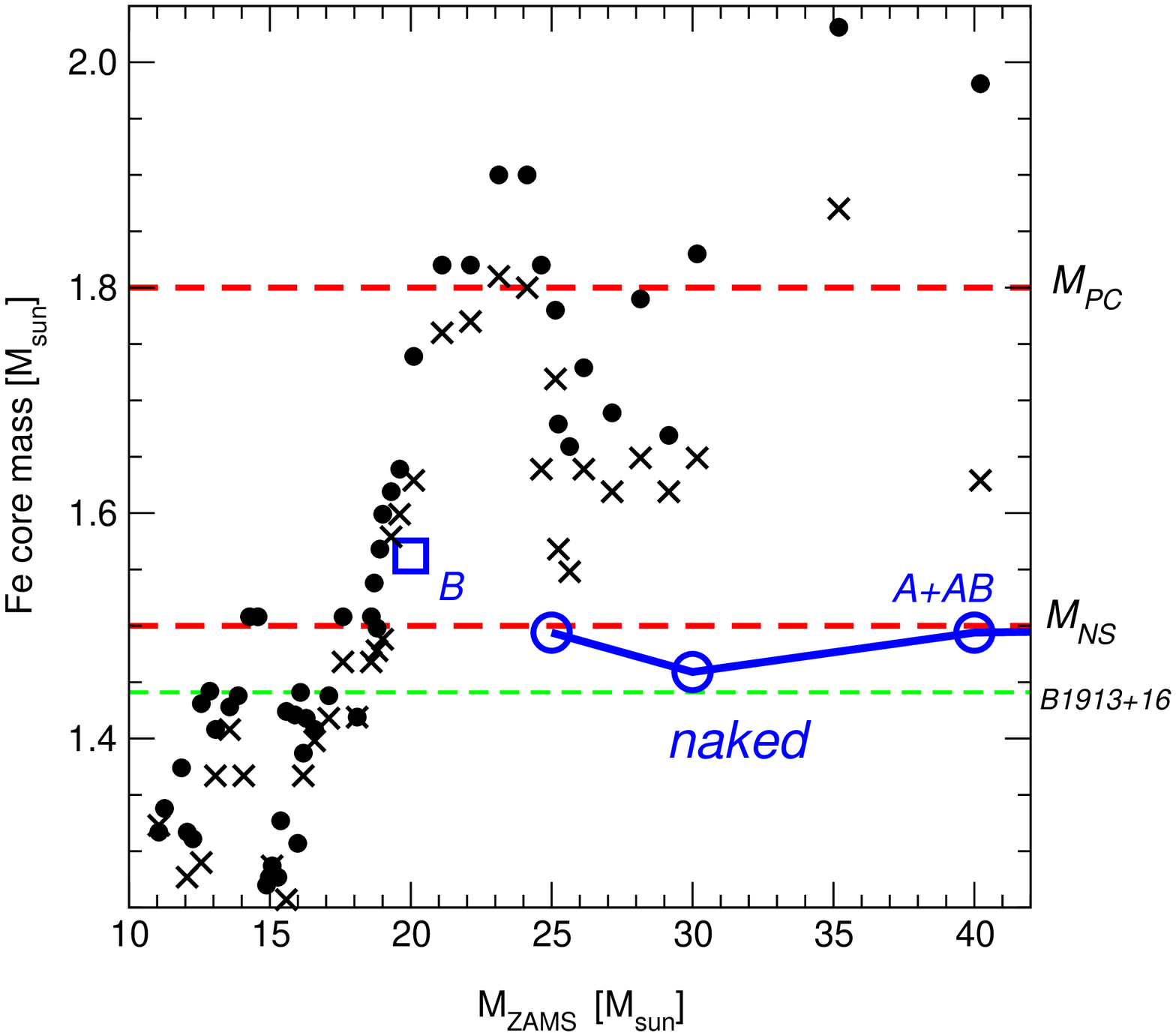}}
\caption{Comparison of the iron core masses at the time of iron core implosion for a finely spaced grid of stellar masses \citep{HWMPL01}.  The circular black dots were calculated with the \citet{Wos95} code, whereas the crosses employ the vastly improved \citet{Lan00} rates for electron capture and beta decay.  If the assembled core mass is greater than $M_{PC}=1.8\msun$, where $M_{PC}$ is the proto-compact star mass as defined by \citet{BB94}, there is no stability and no bounce; the core collapses into a high mass BH.  $M_{NS}=1.5\msun$ denotes the maximum mass of NS \citep{BB94}.  The mass of the heaviest known well-measured pulsar, PSR 1913$+$16, is also indicated with dashed horizontal line \citep{Tho99}.}
\label{fig:iron}
\end{figure}

Given the Woosley value for the $^{12}$C$+\alpha\rightarrow^{16}$O process of $170$keV barns (at $E=100$keV), to be correct, the best experiments \citep{Kunz} obtaining $165\pm 50$keV barns, the $^{12}$C abundance drops below $15\%$ just around $18\msun$, the mass of the progenitor of SN1987a, Sanduleak $69^\circ$202.  We reproduce Fig.1 of \citet{Bro01} as our Fig~\ref{fig:carbon}.  Thus with Case C mass transfer the threshold for black hole production is ZAMS $18\msun$.  We show the calculated compact object masses from \citet{Bro01} as Fig~\ref{fig:iron}.  This is independent of metallicity, depending only on the low-energy burning.  In fact, the metallicity is essentially zero at the threshold in ZAMS masses for black hole formation.

\citet{Fry04} and \citet{Fry06} independently obtained a ZAMS mass of $\sim20\msun$ to give the lowest mass black hole.  Our use of the Woosley Ansatz is not only a simple, elegant argument, based on the behavior of entropy, but also a practical one.  For example, it settles the questions of central engine in GRB060218/SN2006aj, which was debated by the 119 observers who wrote 5 papers: \citet{N0}, \citet{N1}, \citet{N2}, \citet{N3} and \citet{N4}, mostly speculating that the central engine was a magnetar, although they gave no mechanism by which it could unwind its magnetic field to power the GRB.  If one looks at the additional information of \citet{N4}, one sees that there were no carbon lines.  (Actually, these lines need not be zero, but some weak ones can come from carbon shell burning.)  Therefore, the explosion was I$_{bcd}$ in nature.  This occurs only in one venue in astronomy; namely, in a black hole formed at the lowest ZAMS mass possible, $\sim18\msun$.  Thus, clearly the central engine was a rotating black hole.  The hypernova had a low energy of $2$ bethes \citep{N4} and the GRB was highly subluminous.  However, from equipartition of energy we expect the kinetic energy to be roughly equal to the thermal energy, so the kinetic energy must have been mostly used up by the ram pressure which cleared the way for the GRB (For GRB 980425, \citet{MFThesis} estimates this to be $\sim10^{52}$ergs). Clearly the ``unwinding" of the lowest possible rotational energy will continue until the GRB falls below observational threshold.  The $T90$ of GRB 060218 is $2100\pm100$s \citep{N1}, extremely long.  It is clear that if all of the rotational energy can be accepted, which should certainly be true for a total energy of only $E_{rot}\sim4\times10^{51}$ergs, then the time will be very long because the kinetic energy is mostly used up in clearing the way through the star for the GRB, which has power $\propto B^2$, where $B$ is the magnetic field, which would be expected to be low for the lowest mass black hole.  We understand in this way that the subluminous bursts from low rotational energy have a long lifetime.  \citet{Kan07} find the energies, kinetic and thermal, to be the smallest in GRB 060218 among the subluminous bursts, GRBs 980425, 030329, 031203 and 060218, they investigate.  It should be noted that the very low GRB energy comes from a near cancellation of the kinetic energy by the ram pressure work.  Note that with such an energy this would be a nearly ``dark" explosion like Cyg X$-$1 \citep{BLMM07}, which was estimated to have $\sim6$ bethes of rotational energy, except that the black hole must be $\gtrsim1.5\msun$ at the time of explosion rather than the $\sim7\msun$ black hole calculated by \citet{BLMM07} for Cyg X$-$1.  In the case of Cyg X$-$1, extensive mass transfer from the donor to the black hole and mass loss from the donor complicated the simple interpretation using Kepler's law.  None the less, the $\sim6$ bethes obtained from Cyg X$-$1 should not be so different from the $\sim4$ bethes for GRS 060218 because the donor in the latter must be a factor of $\sim3$ lower in mass than the estimated $\sim30\msun$ in Cyg X$-$1, in order to be less massive than the black hole progenitor.  This would increase the rotational energy by a factor of $\sim3$ whereas the black hole in GRS 060218 is $\gtrsim 1.5\msun$, a factor of $\sim14/3$ less than that of the $7\msun$ black hole in Cyg X$-$1 at the time of common envelope evolution.  Scaling 6 bethes by $3/(14/3)$ gives $\sim 4$ bethes.  \citet{Mirabel} found the mass loss in the explosion of Cyg X$-$1 to be very small, the space velocity of Cyg X$-$1 relative to the Cyg OB3 association to be typical of the velocities of stars in expanding O-star associations and gave other arguments supporting a ``dark" explosion.  The truly remarkable property of GRB 060218 is the long $T_{90}$ time of the GRB of $2000\pm100$ seconds. This is, however, the typical time of the Blandford-Znajek engine (eq.(8) of \citet{Lee00})
\be
\tau_{BZ}=\frac{M_{BH}c^2}{B^2R^2c^2} \sim2.7\times10^3 \left(\frac{10^{15}G}{B}\right)^2 \left(\frac{\msun}{M}\right){\rm sec}.
\ee
The low magnetic field, appropriate for low-mass black holes and the low $M_{BH}$ tend to make $\tau_{BZ}$ longer, such that the measured $T_{90}=2000\pm100$ sec. is just the right time for the GRB in 060218.
That of GRB 060218 is certainly no more than ``dusky".

We feel confident that the subluminous GRBs can be described by our binary scenario.  As noted, we do not claim that the cosmological ones can be; they may require the original Woosley model.  However, our model is not really different, in the sense that we use the Woosley collapsar model after the donor has spun up the black hole progenitor to the necessary amount of rotation, and we feel that understanding the dynamics of LMC X$-$3 will help us make a connection between the subluminous and cosmological GRBs.

%--------------------------------------------------------------------
\section{Why Case C Mass Transfer?}\label{CaseC}

It is clear that Case C, mass transfer following He burning, is useful in the binary evolution of black holes, especially in a Galaxy with solar metallicity.  In this case the large winds off the stellar surfaces, referred to by \citet{Fruchter06}, do not blow away the He because it is covered by hydrogen -just like in single stars, and, therefore, protected from the strong He-star winds.  Were this not so, the winds in our solar-metallicity Galaxy, {\it would} blow off the helium, lowering the stellar masses sufficiently that they would go into neutron stars, the fate described by \citet{Fruchter06}.  The idea that life in the Milky Way is protected by the metals keeping the GRBs away \citep{Sta06} obtained from \citet{Fruchter06} is correct in that they do keep the high-luminosity long GRBs away.  The idea that the black hole progenitor does not tidally lock with the donor, until the explosion, found in the literature of the soft X-ray transient sources is incorrect, as we now discuss.

The tidal locking between the donor and the He star has been demonstrated by \citet{vdH07}.  Given this, when the center of the He star falls into the preforming black hole, the metals to which it has burned before Case C mass transfer undergo a perfect magnetohydrodynamics.  The magnetic field is frozen in the material, and transfers the angular momentum of the He star to the black hole as it forms.  The high, $\sim10^{15}$G, B-field threading the disk of the black hole transfers the tidal locking of the He star with the donor to that of the black hole and the donor.  Then the donor decouples and takes the role of an observer to the explosion.  As noted earlier, by the material the donor accretes the nature of the explosion can be reconstructed.

Aside from the help in tidal locking by the metals threading the disk, the Case C mass transfer accomplishes two other mechanisms.  Firstly, the remaining helium is originally on the outer part of the He star, later supported by the angular momentum that cannot be transferred into the black hole, so that it can leave in the Blaauw-Boersma explosion without interaction, thereby leaving a type I$_c$ explosion.  There is then no helium envelope to tie the magnetic fields to, which would tend to reduce the angular momentum of black hole \citep{HWLS}.  As can be seen from \citet{Lee02}, Case C mass transfer also makes the initial binary separation $a_i$  insensitive to donor mass, making our use of Kepler's law easy.

This latter argument results from the nature of the He supergiant at the time of mass transfer \citep{Lee02}.  In order for the mass transfer to be delayed sufficiently for Case C mass transfer to take place, winds must be sufficiently low that the metals obtained at the end of the burning do not blow away, but extend in space beyond the helium.  Otherwise Case B mass transfer would take place.  For a ZAMS $20 \msun$ giant, this means that the He supergiant must expand to $\sim1000 \rsun$.
Adding the $\sim(200-300) \rsun$ Roche lobe of the donor gives mass transfer beginning at $(1200-1300) \rsun$.

\begin{figure}
\centerline{\includegraphics[width=0.5\textwidth]{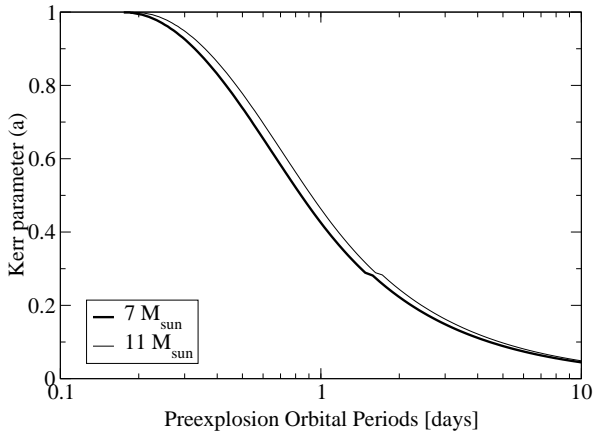}}
\caption{The Kerr parameter of the black hole resulting from the collapse of a helium star synchronous with the orbit, as a function of orbital period (LBW). Note that the result depends very little on the mass of the helium star, or on whether we use a simple polytrope or a more sophisticated model.}
\label{fig:kerr}
\end{figure}

In either case, the mass transfer is well localized in radius because the donors will have sizes which are small compared with the radii of the supergiants, so that their Roche lobes are typically about $0.2$ of that radius.  In \citet{Lee02} we found all supergiants to have $\sim 30\msun$.  This was because in the Galactic evolution, binaries with lesser mass black holes would lose more than half their mass in the explosion and not be stable \citep{Mor07}, but with the lower rotational energies of the low metallicity binaries, the lower-mass giants of $\sim 20\msun$ are likely to be more copious.  These would allow an $a_i\sim(1200-1300)\rsun$ for Case C mass transfer, without strong variation, so it is reasonable to forget the binary dependence of $a_i$.  If we can do this we can easily use eq.~(4) of \citet{Mor07}
\be
\frac{\rm days}{P_b}= \left(\frac{4.2\rsun/a_i}{M_d/\msun}\right)^{3/2} \left(\frac{M_d+M_{He}}{\msun}\right)^{1/2} \left(\frac{M_{giant}}{\msun}\right)^{0.83}, \label{eq:afKepler}
\ee
where it is seen that the approximation
\be
\frac{\rm days}{P_b}\propto \left(\frac{4.2\rsun/a_i}{M_d/\msun}\right)^{3/2} \left(\frac{M_d+M_{He}}{\msun}\right)^{1/2} \label{eq:afKeplerProp}
\ee
gives the main dependence as
\be
\frac{P_b}{days}\propto\left(\frac{M_d}{\msun}\right)^{3/2}
\ee
for $M_d \ll M_{He}$ and
\be
\frac{P_b}{days}\propto\frac{M_d}{\msun}
\ee
for higher $M_d$.  These can be normalized to Galactic binaries, say Nova Sco or Il Lupi, where the theoretical and observational Kerr parameters agree.  We reproduce Fig.1 of \citet{BLMM07} as our Fig.~\ref{fig:kerr} to show how it is possible to obtain the Kerr parameter from $P_b$.

%--------------------------------------------------------------------
\section{LMC X$-$3}\label{LMCX3}

The poster child subluminous explosion GRB 980425/SN1998bw came from an LMC-size star-forming Galaxy but with metallicity nearly solar \citep{Sollerman05}.

The LMC has $\sim1/3$ solar metallicity, so it should go a long way towards having low metallicity stars.  The donor of LMC X$-$1 is more massive than the black hole and it seems to be somewhere between Cyg X$-$1 and M33 X$-$7 in mass and their relative sizes.
In the case of LMC X-3
\citet{Davis06} find $a_\star\simeq0.26$ now; using the \citet{Cow83} $7\msun$ for the black hole would imply using $4\msun$ for the donor.  The present period is $1.7$days, with present rotational energy of $54$ bethes, obtained from the period from Fig.~\ref{fig:kerr}.

Note that the much lower energy than those of most Galactic binaries result from the $2-4$ times larger donor mass here.  It would be natural for the natal rotational energy to be about double this, with energy $\sim108$ bethes.  The hypernova energy of SN1998bw is $\sim30$ bethes, and, as noted earlier, the kinetic energy should be about equal to this because of equipartition of kinetic and thermal energies.  The GRB 980425 is a ``smothered" one \citep{MFThesis}, but, in general, the visible GRB is only $\sim1$ bethe, so in most cases the energy used up by the ram pressure in order to clear out material from the path of the jet must be cancelled in first order of magnitude by the kinetic energy.
We believe this near cancellation between kinetic energy and energy needed for the ram pressure work to produce the various observed low-energy GRBs (see our later discussion).

We believe that adding the $\sim30$ bethes kinetic energy to the $30$ bethes measured hypernova explosion energy, assuming equipartition between the two energies, gives us an energy of $\sim60$ bethes, possibly that of cosmological GRBs.  At least, this is the largest energy that we have found recently in the literature.  We argue that GRB 980425 is subluminous because of the near cancellation between work of the ram pressure and the kinetic energy.  (But it could be subluminous because of the high, nearly solar metallicity in the background galaxy.)  %In any case, %measurements now proceeding by the Smithsonian-Harvard group should give us more information.

The present rotational energy of LMC X$-$3 is $\sim54$ bethes, obtained from the $1.7$ day period and Fig.~\ref{fig:kerr}, roughly half of the natal rotational energy if we assume the explosion to use up $\sim60$ bethes. Going back to Fig.~\ref{fig:kerr} again, this would imply a pre-explosion period of $\sim1$ day for LMC X$-$3.

It is clear that the black-hole binaries cannot deliver all of their rotational energy into the explosion because their power goes to zero as they slow down.  From dimensional analysis it would seem reasonable that their pre-explosion energy would be about double the post-explosion energy, assuming that all of this latter energy could be accepted.  In any case, our assumption can be tested in that the system velocity, at least the radial component of it, will be measured by the Smithsonian-Harvard collaboration.  In our above model we have a mass loss of $\Delta M = 3.25\msun$, which implies a system velocity of $43$ km/s, definitely not a ``dark" explosion.  This is a little less than $1/3$ the system velocity of Nova Sco, where the peculiar velocity in the direction of the local standard of rest is $(114\pm19)$ km/s \citep{Brandt95}, where only one component of Nova Sco's space velocity has been measured, so that its space velocity may be higher.  In any case, it is clear that LMC X$-$3 is an interesting black-hole binary, and better measurement of its properties is underway.

We note that LMC X$-$1 has a donor mass more massive than its black hole \citep{Orosz02}, ${M_{BH}}/{M_{donor}}\sim(0.3-0.7)$ with black hole mass $\sim(7-13)\msun$.  Thus LMC X$-$1 is similar to Cyg X$-$1, and  would have been expected to undergo a ``dark" explosion.

%--------------------------------------------------------------------
\section{Conclusions}\label{Conclusion}

We suggest that LMC X-3 may be similar to relics of cosmological GRBs, to the extent that some or most of these latter result from binaries.  The number of the latter is certainly sufficient to produce the GRBs and the binary nature takes care of the necessary angular momentum, which can be achieved by choosing the donor mass.

\citet{Fruchter06} have made a case that high-luminosity, long GRBs came from irregular low-metal galaxies.  We suggest LMC X-3 as the closest nearby binary from a region of $1/3$ solar metallicity in an irregular galaxy.  The lower metallicity environment has more massive stars, and the donor in LMC X-3 is probably at least twice as massive as the donors of the relics of the Galactic GRBs, which should slow the binary down to a rotational energy that can be accepted.  We believe that LMC X-3 brings us in the metallicity towards the irregular low-metallicity binaries considered by \citet{Fruchter06}.

Working in this region of energies we feel that we can make predictions, because our calculation of Kerr parameters makes it possible to make quantitative calculations.

Our predictions are, however, at best, order of magnitude, because the necessary properties of the binary LMC X-3 have not been accurately measured.  As they are measured we will probably have to readjust our numbers.  In particular, it would be most valuable to obtain a reliable value for the explosion energy.  Whereas the hypernova energy is probably measured with reasonable accuracy, the kinetic energy required by the work performed by the ram pressure to clear the way for the GRB is mostly hidden, and the net GRB energy is a small difference between large kinetic energy and a large amount of work done by the ram pressure.  We believe that there will be great variations in GRBs depending upon these small differences which must be sensitive to stellar properties.

Our present estimate of $\sim 60$ bethes for the total cosological GRB energy is the highest recent estimate that we have seen.  For this estimate we invoke equipartition of kinetic and thermal energies.

Finally, we wish to point out that our ability to calculate Kerr parameters and the confirmation of these by the Smithsonian-Harvard measurements makes it possible to calculate in a reliable way the amount of angular momentum carried by the black-hole binary.  This angular momentum is conserved, but the questions to be answered concern how it is to be distributed:  how much remains in the binary, how much energy goes into the explosion and what is the fraction of the latter which goes into kinetic and into heat energy?  Progress in answering these questions is necessary to make a quantitative study out of the GRBs, and to put order into their classification.

%--------------------------------------------------------------------
\section*{Acknowledgments}

We would like to thank Jeff McClintock for many useful discussions.
G.E.B. was supported by the US Department of Energy under Grant No. DE-FG02-88ER40388.
CHL was supported by Creative Research Initiatives (MEMS Space Telescope) of MOST/KOSEF.

\appendix
%--------------------------------------------------------------------
\section{Appendix A:  Energies}\label{A1}

The energies of cosmological GRBs and hypernovae are generally taken to be $\sim1$ bethe and a few times $10$ bethes, respectively.  However, the GRB involves, in order that it can begin, the clearing of the matter in the way of the jet in order that it can leave the star.  This energy is roughly the mass still contained within the beaming angle of the jet times the square of the velocity with which it is displaced.  This is about $1\msun$ (including both poles) times $(1\%-10\%) c^2$ \citep{MFThesis}.  MacFadyen estimates this to be a few $\times10^{51}$ ergs.  However, the$(1\%-10\%)\msun c^2$ is $\sim(2\times10^{52}-2\times10^{53})$ ergs.  We compare this with the thermal energy of $\sim3\times10^{53}$ ergs in the hypernova 1998bw.  The mechanism for heat production is viscous heating, from the field lines frozen in the rotating disk, different from ordinary supernovae.  In the latter, the kinetic and heat energy are roughly equal, from energy equipartition.  We assume this to be true also in the GRB-Hypernova case.  Otherwise we cannot make sense out of our energies.  In other words, we believe that GRB 980425/SN 1998bw had essentially the same rotational energy as cosmological GRBs, the GRB, however, being nearly smothered by the work from the ram pressure.

This means that the kinetic energy is roughly equal to that of the hypernova, $\sim30$ bethes in the case of GRB 980425.  The GRB is, however, a ``smothered" one \citep{MFThesis}, which means that the ram pressure to free the jet uses up nearly all of the kinetic energy.

In cosmological GRBs the typical GRB energy is $\sim1$ bethe, resulting from a near cancellation of work from ram pressure work and other kinetic energy.

If such a near cancellation is to be engineered, the helium stars in which the jet is produced cannot be very different in properties.  However, the cosmological GRBs are only a tiny population compared with subluminal ones, which are probably formed with helium stars that do not show this near cancellation of kinetic energies with ram pressure work.

To summarize, we are saying that there is a large ``invisible energy"  which goes into work done by the ram pressure, so that our energies calculated in the Blandford-Znajek mechanism look much larger than those found empirically in the observations, although it is generally realized that extra energy must be furnished to get photons out in the GRB, by saying that there is an efficiency, often taken to be $\sim 10\%$.

\section{Appendix B: A New Constraint for GRB Progenitor Mass }

We wish to point out that there is observational support for our choice of the LMC as a site for the long gamma-ray bursts. \citet{Lar07} provide a new constraint for gamma-ray burst progenitor mass. They show that long-duration gamma-ray bursts (L-GRBs) are much more concentrated
on their host galaxy light than core collapse supernova explosions. From this they way {\it ``Assuming core collapse supernova arise from stars with main-sequence masses $>8\msun$, GRBs are likely to arise from stars with initial masses $>20\msun$. This difference can naturally be explained by the requirement that stars which create a L-GRB must also create a black hole."}

As a template and close by natural analogue of starburst galaxies they use NGC 4038/39. They say, however, {\it the luminosity function and surface density of clusters on NGC 4038/39 is comparable to that seen in other local star-forming galaxies of varying morphology in which they include the LMC.

%--------------------------------------------------------------------

%%%%%%%%%%%%%%%%%%%%%%%%%%%%%%%%%%%%%%%%%%%%%%%%%%%%%%%%%%%%%%%%%%%%%%%%%%%%%%%%

\end{document}